\documentclass[aip,apl,amsmath,amssymb,amsfonts,reprint]{revtex4-2}

\usepackage{dcolumn}
\usepackage{siunitx}
\usepackage{easyReview}

\usepackage[normalem]{ulem}
\usepackage{bm} 
\usepackage{amssymb}
\usepackage{amsfonts} 
\usepackage{amsmath} 

\usepackage{graphicx} 

\usepackage{xcolor} 

\usepackage[pdftex,colorlinks=true,allcolors=blue,breaklinks=true]{hyperref}  

\usepackage[utf8]{inputenc} 
\usepackage{textgreek}

\definecolor{g-blue}{rgb}{0.83,0.95,1}
\definecolor{g-yellow}{rgb}{1,1,0.7}
\definecolor{g-green}{rgb}{0.9,1,0.9}
\definecolor{green}{rgb}{0,0.6,0}
\definecolor{cyan}{rgb}{0,0.7,0.7}
\definecolor{black}{rgb}{0,0,0}
\definecolor{grey}{rgb}{0.4,0.4,0.4}
\definecolor{nature-blue}{rgb}{0.0,0.200,0.500}

\def \ed {\end{document}}
\def\Fbox#1{\vskip1ex\hbox to 8.5cm{\hfil\fboxsep0.3cm\fbox{%
		\parbox{8.0cm}{#1}}\hfil}\vskip1ex\noindent}  

\def\be{\begin{equation}}
\def\ee{\end{equation}}
\def\bea{\begin{eqnarray}}
\def\eea{\end{eqnarray}}
\def\bse{\begin{subequations}}
\def\ese{\end{subequations}}

\let \= \equiv  
\let\*\cdot 
\let\^\widehat 
\let\-\overline

\def\1{\bm1}

\def\<{\left\langle}    \def\>{\right\rangle}
\def\({\left(}          \def\){\right)}
\def\[ {\left[}         \def\]{\right]}

\newcommand{\Eq}[1]{Eq.\,(\ref{#1})}
\newcommand{\Eqs}[1]{Eqs.\,(\ref{#1})}
\newcommand{\Fig}[1]{Fig.\,\ref{#1}}
\newcommand{\Figs}[1]{Figs.\,\ref{#1}}

\newcommand{\B}[1]{{\bm{#1}}}

\renewcommand{\sb}[1]{_{\text {#1}}}  
\renewcommand{\sp}[1]{^{\text {#1}}}  

\newcommand{\RPTU}{\affiliation{Fachbereich Physik and Landesforschungszentrum OPTIMAS, Rheinland-Pf\"alzische Technische Universit\"at Kaiserslautern-Landau, 67663 Kaiserslautern, Germany}}
\newcommand{\Fraunhofer}{\affiliation{Fraunhofer Institute for Industrial Mathematics ITWM, Fraunhofer-Platz 1, 67663 Kaiserslautern, Germany}}
\newcommand{\WeizmannChemicalBiologicalPhysics}{\affiliation{Department of Chemical and Biological Physics, Weizmann Institute of Science, Rehovot 76100, Israel}}
\newcommand{\WeizmannComplexSystems}{\affiliation{Department of Complex Systems, Weizmann Institute of Science, Rehovot 76100, Israel}}

\begin{document}
\title[Applied Physics Letters---Special Topic on Magnonic]{Local temperature control of magnon frequency and direction \\ of supercurrents in a magnon Bose--Einstein condensate}

    \author{Matthias R. Schweizer}
    \email{mschweiz@rptu.de}
    \RPTU
    
    \author{Franziska Kühn}
    \RPTU

	\author{Victor~S.~L'vov}
    \WeizmannComplexSystems
    \WeizmannChemicalBiologicalPhysics
	
	\author{Anna Pomyalov}
	\WeizmannChemicalBiologicalPhysics
	 
	\author{Georg~von~Freymann}
	\RPTU
    \Fraunhofer
    
    \author{Burkard~Hillebrands}
	\RPTU

	\author{Alexander~A.~Serga}
	\email{serha@rptu.de}
	\RPTU

\date{\today}

\begin{abstract}  
The creation of temperature variations in magnetization, and hence in the frequencies of the magnon spectrum in laser-heated regions of magnetic films, is an important method for studying Bose--Einstein condensation of magnons, magnon supercurrents, Bogoliubov waves, and similar phenomena. 
In our study, we demonstrate analytically, numerically, and experimentally that, in addition to the magnetization variations, it is necessary to consider the {connected} variations of {the} demagnetizing field{. In case of a heat induced local minimum of the saturation magnetization, t}he combination of these two effects results in a local increase in the minimum {frequency }value of the magnon {dispersion} at which the Bose--Einstein condensate emerges. As a result, a magnon supercurrent directed away from the hot region is formed.

\end{abstract}

\maketitle 

The phenomenon of Bose--Einstein condensation, predicted by Einstein\cite{Einstein1925in2005} for an ideal gas and {subsequently} by Fr\"{o}hlich\cite{Froehlich1968} for quanta of collective excitations, has been attracting the attention of the scientific community for a long time. 
Such attention is warranted not only by the universality and physical depth of this phenomenon but also by such practically significant consequences as coherency, superfluidity, and superconductivity. In overpopulated gases of excitons,\cite{Eisenstein2004} magnons,\cite{Bunkov2008, Demokritov2006} photons,\cite{Klaers2010} polaritons\cite{Amo2009}, etc. the Bose--Einstein condensate (BEC) manifests itself as a spontaneous occurrence of macroscopic coherent oscillations at the lowest frequency of the spectrum.\cite{Snoke2006} 
For magnons in a magnetic insulator such as yttrium iron garnet (YIG),\cite{Cherepanov1993, Arsad2023} this condensation can be achieved even at room temperature,\cite{Demokritov2006} which is {relevant} {for} practical applications.\cite{Dzyapko2008, Nakata2015, Pirro2021, Schneider2021_bullet, Mohseni2022, Breitbach2023}
The same applies to magnon supercurrents---a collective motion of condensed magnons driven by a phase gradient $\nabla \varphi$ of the BEC wave function $\psi(x,t)$.\cite{Bozhko2016, Bozhko2019} 
At present, the dynamics of magnon condensates and supercurrents remains intriguing and not fully understood as it can be affected by the spatial distribution of magnetization $M$\cite{Bozhko2016, Bozhko2019, Schweizer2022}, variations in the bias magnetic field $H\sb{ext}$\cite{Kreil2021}, and by various nonlinear effects\cite{Dzyapko2017, Borisenko2020}.

Here, {for a tangentially magnetized magnetic film,} we show theoretically and experimentally that a spatially localized decrease in {the }saturation magnetization {$M_\mathrm{s}(\B{r}, T) $} induced by {local }optical heating leads to a local increase {of} the BEC frequency, which is consistent with the observed supercurrent direction and opens new possibilities for controlling transport in magnon condensates.{ As discussed in the following, this effect is caused by the demagnetizing field generated by the local variations in $\B{M}(\B{r}, T) $.}



\begin{figure} 
 \includegraphics[width=1\columnwidth]{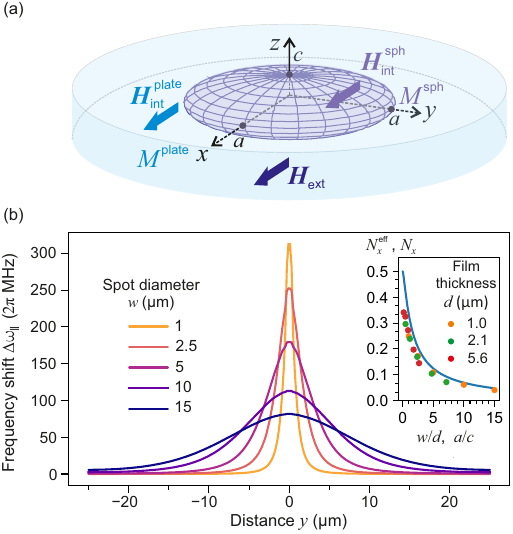}
   \caption{\label{f:model} 
   (a) A spheroidal magnetic inhomogeneity embedded in a tangentially magnetized unbounded magnetic plate. 
   (b) Frequency shift $\Delta \omega_\parallel (y)$ calculated from \Eqs{3} using the internal magnetic field  $H\sb{int}$ obtained from micromagnetic simulations for different hot-spot sizes and fixed $\Delta M$ in a $\SI{2.1}{\micro\meter}$-thick YIG film.
        Inset shows the effective demagnetization factor $N_x^\mathrm{eff}(w/d)$, \Eq{6}, as a function of the ratio of the spot diameter $w$ to the thickness of the plane $d$. The solid blue line shows the analytical dependence of $N_x$ in the spheroid (a) on the ratio of its axes $a/c$\,.
           }
\end{figure}

To provide a qualitative description of the expected phenomena, let us consider the geometry, shown in \Fig{f:model}(a). It consists of an unbounded {magnetic} plate with the surface in the plane $(\widehat{\B{x}},\widehat{\B{y}})$ placed in a tangential magnetic field $\B H\sb{ext}$, which is aligned with the $\widehat{\B x}$ axis. The magnetization $\B M\sp{plate}$ is also parallel to $\widehat {\B x}$.
{For simplicity, let us assume that the} plate hosts an ellipsoid of revolution (spheroid) around $\widehat{\B z}$, whose axis ratio $R = c/a$ is a parameter of the problem. The magnetization of this spheroid $M\sp{sph}{(T)}$ is smaller than $M\sp{plate}$. To find the intrinsic magnetic fields in the plate and in the spheroid, $\B H\sb{int}\sp{plate}$ and $\B H\sb{int}\sp{sph}$, respectively, we use
the continuity condition across a surface for the orthogonal component of the magnetic flux density $\B B$. We obtain that in the plate $\B H\sb{int}\sp{plate} = \B H\sb{ext}$.

The situation in the spheroid is more involved. We know that the intrinsic magnetic field $\B H\sb{int}\sp{sph}$ in the ellipsoid, placed in the homogeneous external magnetic field $\B H\sb{ext}$, is also homogeneous. When $\B H\sb{ext}$ is oriented along one of the ellipsoid axes ($\widehat{\B x}$, $\widehat{\B y}$ or $\widehat{\B z}$) the intrinsic field is also oriented along this axis (say, $\widehat{\B x}$). 
Assuming that there is no magnetization around the ellipsoid, i.e., $M\sp{plate}=0$, we find the following equation: 
\begin{subequations}\label{1}
  \begin{equation}\label{1a}
    H\sb{int} \sp{sph}=H\sb{ext}- 4\pi N_x M\sp{sph}\ .
  \end{equation}
Here, $N_x$ is the demagnetization factor that varies with the value of $R$. 
The three factors $N_j$ ($j=x,y,z$) satisfy the sum rule $N_x+N_y+N_z=1$. 
In our geometry, we conclude that $N_x$ varies between 1/2 and 0 [see inset in \Fig{1}(b)].\cite{Demagnetization_factors}

When the spheroid is surrounded by a medium with magnetization $M\sp{plate} \ne 0$ (as in our case), the condition of continuity of the tangential component of the magnetic flux $(-M\sp{sph})$ in \Eq{1a} is replaced by $M\sp{plate}-M\sp{sph}$. Thus, we have
  \begin{equation}\label{1b}
    H\sb{int} \sp{sph}=H\sb{ext} + 4\pi N_x \Delta M, \ \Delta M=M\sp{plate} - M\sp{sph} \ .
  \end{equation}
\end{subequations} 
As expected, when $M\sp{plate} = M\sp{sph}$, the magnetic field inside the spheroid {is }$H\sb{int} \sp{sph}=H\sb{ext}$, and when $M\sp{plate}=0$ the intrinsic field $H\sb{int} \sp{sph}$ is determined by \Eq{1a}.
 
Consider now how the variations in magnetization affect magnon frequencies $\omega(\B k)$. For exchange magnons {we find} (see Eq.\,(7.9) in Ref.\,\onlinecite{Gurevich-Melkov1996})
\begin{align}
  \begin{split}\label{2} 
    \omega (\B k) =& \big[\omega_{_H} + \eta k^2\big]^\frac{1}{2} \\ 
    &\times \big[\omega_{_H} + \eta k^2 + \omega_{_M} \sin ^2 \theta\big]^\frac{1}{2}  \\
    \omega_{_H}=& \gamma(H\sb{ext}-4\pi N_x M)\,, \quad  \omega_{_M}= \gamma\, 4\pi M\ .
  \end{split}   
\end{align} 
Here, $\eta$ is the nonuniform exchange constant, $M$ is the magnetization of the medium, $\theta$ is the angle between $\B k$ and $\B M$, and $\gamma$ is the gyromagnetic ratio.

Below, we describe the effect of temperature modification of {the} magnon {spectrum} {using} the example of two typical frequencies with $\theta=0$ and $\theta=\pi/2$ and $k \to 0$. These frequencies are easily available for experimental study. 

 In the limit  $k\to 0$ and $\B k\|\B M$, in a tangentially magnetized plate (where $H\sb{int}=H\sb{ext}$)
\begin{subequations}\label{3}  
  \begin{equation}\label{3a}
    \omega_\|\sp{plate} = \gamma\, H\sb{ext} \,,
  \end{equation}
while in the spheroid,  where $H\sb{int}$ is given by \Eq{1b}, 
  \begin{equation}\label{3b}
    \omega_\|\sp{sph} = \gamma (H\sb{ext} + 4\pi \Delta M N_x) = \omega_\|\sp{plate} + \gamma  4\pi \Delta M  N_x \ .
  \end{equation}
\end{subequations} 
We see that{,} when $M\sp{plate}>M\sp{sph}$, the bottom magnon frequency in 
the spheroid is larger than this frequency in the surrounding plate. 

The frequency of magnons with $\B k \perp \B M$ (i.e. for $\theta =\pi/2$) and with $k \to 0$ also follows from \Eq{2}.
\begin{subequations}\label{4} 
Thus, in the plate, we have
  \begin{equation}\label{4a}
    \omega_\perp\sp{plate} = \gamma \sqrt{H\sb{ext}(H\sb{ext}+4\pi M\sp{plate})} \,,
  \end{equation} 
while in the spheroid
  \begin{align}\label{4b}
    \begin{split} 
      \omega_\perp\sp{sph} = &\gamma [ ( H\sb{ext}+4\pi N_x \Delta M) \\  
      &\times ( H\sb{ext}+4\pi N_x \Delta M + 4 \pi M\sp{sph}) ]^{1/2} \ .
    \end{split}
  \end{align}  
\end{subequations} 

Comparing \Eqs{4a} and \eqref{4b} and assuming for simplicity that $\Delta M\ll M\sp{plate}$ we see that $\omega_\perp\sp{sph} > \omega_\perp\sp{plate}$ for a prolate spheroid with $N_x=1/2$, while $\omega_\perp\sp{sph} < \omega_\perp\sp{plate}$ for an oblate spheroid with $N_x=0$. The critical value of $N_x$ at which $\omega_\perp\sp{sph} = \omega_\perp\sp{plate}$ is as follows:
\begin{equation}\label{cr}
  N_x\sp{cr}= \frac 12 \Big [1 +\frac {2\pi M\sp{plate}}{H\sb{ext}} \Big]^{-1}\ .
\end{equation}
We see that $N\sp{cr}_x< 0.5${. It} can be {controlled} by the external magnetic field. 
 
Let us examine the thermal modification of the magnon frequencies under more realistic conditions. For this goal, we conduct micro-magnetic simulations of the internal magnetic field in a tangentially magnetized YIG plate for a given magnetization profile using the open-source GPU-based software MuMax 3.10.\cite{Vansteenkiste2014} 
The chosen geometry is determined by a bias magnetic flux density $B$ of 1300\,G ($B \sb{SI units}$=\SI{130}{\milli\tesla}) and a film thickness of \SI{2.1}{\micro\meter}. The spatial distribution of magnetization was chosen as a cylindrical well with a Gaussian profile and depth $\Delta M$. 
Thus, the model accounts for the magnetization gradient in the plane of the plate, assuming uniform magnetization along its thickness. 
The simulation results in a profile $H\sb{int}(x)$ for the whole sample, which can be recalculated into a profile $\omega_\|(x)$ using a relationship $\Delta \omega_\|(x)=\gamma \Delta H\sb{int}(x)$.

The obtained frequency profiles for the magnetization wells of different diameters and fixed depths $4 \pi \Delta M \simeq 400$\,G are shown in \Fig{f:model}(b).\cite{Magnetization_change}  
{Indeed, we see that} the frequency $\omega_\parallel (y)$ increases in the hot-spot region, and this effect becomes more pronounced (even at constant $\Delta M$) as the hot-spot diameter decreases.

Equating the analytical value of $\Delta H\sb{int}\sp{sph}$ $\eqref{1b}$ to the numerical value of $\Delta H\sb {int}(0)$ at the center of the hot spot, we find the effective demagnetization factor $N_x\sp{eff}$ defined by
\begin{equation}\label{6}
4\pi N_x\sp{eff}=\Delta H\sb{int}(0)\big /\Delta M\ . 
\end{equation}
The resulting values of $N_x\sp{eff}$ are shown in the insert {of} \Fig{f:model}(b) by color dots for different ratios of the spot diameter $w$ to the plate thickness $d$. One sees that the numerical dependence $N_x\sp{eff}(w/d)$ is in good quantitative agreement with the analytical dependence $N_x (a/b)$\cite{Osborn1945} shown by the solid line. It means that $H\sb{int}(0)$ at the center of our magnetization profile is well approximated by $H\sb{int} \sp{sph}$ for a spheroid with $\Delta M = \Delta M(0)$ and $a/b = w/d$. This opens the possibility of analytically finding the magnon frequency profiles of hot spots in magnetic films without numerical modeling.

\begin{figure}[t]
\includegraphics[width=1\columnwidth]{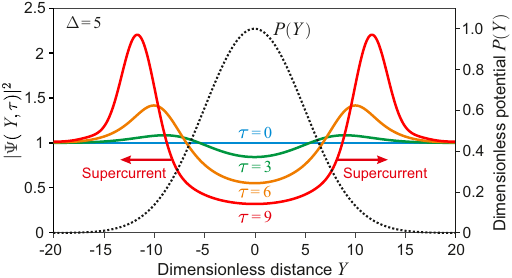}
       \caption{\label{f:GPE_results}
       The BEC wave function $\Psi(Y,\tau)$ \Eq{GPEb} for several moments of the  dimensionless time $\tau$. 
       The potential $P(Y)$ given by \Eq{GPEc} is shown as a dotted black line. 
               }
\end{figure}

To clarify the dynamics of the magnon supercurrent in the vicinity of the hot spot, we numerically solved the Gross--Pitaevskii equation. Since this supercurrent is highest in the direction perpendicular to the magnetic field,\cite{Bozhko2016} we can, in first approximation, limit ourselves to the one-dimensional equation 
\begin{subequations}\label{GPE}
\begin{equation}\label{GPEa}
i \frac{\partial \psi(y,t)}{\partial t}= \Big [ - \frac{\omega ''_{yy} }{2}\frac {\partial ^2}{\partial y^2} + \Omega(x)\Big]\psi(y,t)\,,
\end{equation}
in which $\psi(y,t)$ is the sum of $\psi_+(y,t)$ and $\psi_-(y,t)$, the amplitudes of two BECs\cite{Dzyapko2017, Mohseni2022} with wave vectors $\B k= + \B k_0$ and $\B k= - \B k_0$ in the two frequency minima of the magnon spectrum of the tangentially magnetized magnetic film. The dispersion coefficient $\omega''_{yy}=\partial ^2 \omega (\B k)/(\partial k_y)^2$ calculated at $\B k=\pm \B k_0$ is inversely proportional to the effective mass of condensed magnons. The frequency profile
$\Omega (y) = \Omega\sb{max} \exp( - y^2/(2 \delta^2))$ plays the role of an external potential.

Technically it is more convenient to deal with a dimensionless form of \Eq{GPEa}:
\begin{equation}\label{GPEb}
i \frac{\partial \Psi(Y,\tau)}{\partial \tau}= \Big [ - \frac{1}{2}\frac {\partial ^2}{\partial Y^2} + P(Y)\Big]\Psi(Y,\tau)\ .
\end{equation}  
Here \begin{align}\begin{split}\label{GPEc}
\Psi = & \psi \, \Omega\sb{max} \,, \quad  ~ \tau = t\,\Omega\sb{max}\,, 
\quad Y = y \, \sqrt {\frac {\Omega\sb{max}}{\omega ''}}\,, \\
    \Delta =&  \delta \,\sqrt {\frac {\Omega\sb{max}}{\omega '' }}\,, \quad P(Y)= \exp \Big ( - \frac{Y^2} {2\Delta ^2}\Big )\ .
\end{split}\end{align}\end{subequations}
Equations \eqref{GPEb}-\eqref{GPEc} were solved numerically by the split-step Fourier method.
In \Fig{f:GPE_results}, we present the evolution of $\Psi(Y,\tau)$ from the homogeneous initial condition $\Psi(Y,0)=1$.  

As seen in \Fig{f:GPE_results}, our solution demonstrates a supercurrent propagating outward from the region of the elevated BEC frequency and, hence, decreased magnetization. Further steps towards understanding the magnon supercurrent dynamics require consideration of the nonlinear terms of the 2-dimensional Gross--Pitaevskii equation, in particular those related to the static demagnetization field,\cite{Borisenko2020} which in the case of supercurrents varies with the spatial distribution of the BEC density. This problem is beyond the scope of this paper.

\begin{figure}[tb]
\includegraphics[width=1\columnwidth]{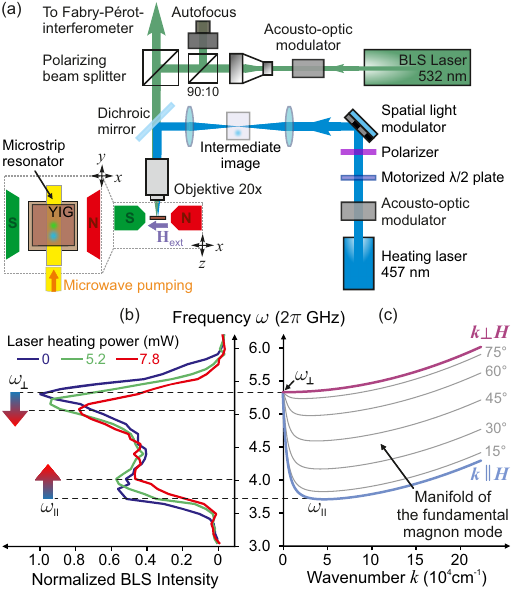}
       \caption{\label{f:setup}(a) Schematic representation of the experimental setup consisting of a BLS spectrometer (green beam path) and heating (blue beam path) module. (b) The thermal magnon spectrum $N(\omega)$ measured for three heating powers at the center of the hot spot. Both a shift of the lower peak of the magnon density at $\omega_\|$ to higher frequencies and a shift of the upper peak at $\omega_\perp$ to lower frequencies are visible. (c) The dispersion curves for the dipole-exchange fundamental magnon mode in a tangentially magnetized YIG film are shown for different angles $\theta$ between $\B k$ and $\B M$.}
\end{figure} 

\begin{figure*}[t]
\includegraphics[width=2\columnwidth]{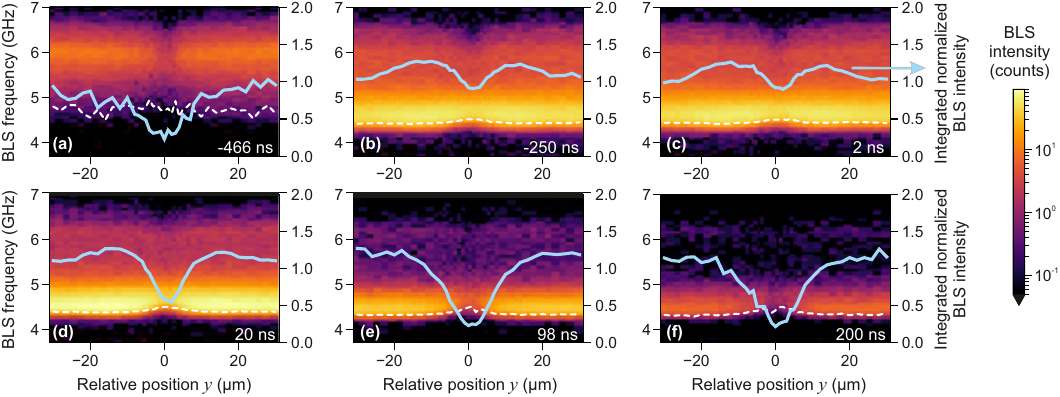}
       \caption{\label{f:freq_intens} Overview of the development of the magnon spectrum. (a) Just after the start of the parametric amplification. (b) During the pumping pulse in a stable situation. (c) and (d) after pumping has been switched off at $t=0$. (e) and (f) long after pumping has been switched off. The dashed lines mark the lower population frequency of the magnon spectrum, and the solid blue lines represent the BLS intensity integrated in the range from 4\,GHz to 5.2\,GHz and normalized to the reference signal without heating. The frequency resolution is limited by the interferometer line width and is about 100\,MHz.}
\end{figure*} 

In order to experimentally verify the theoretical results, we used a dedicated optical system [see \Fig{f:setup}(a)] that allows us to create thermal profiles of various forms and various sizes, in particular, to produce a hot spot with a diameter down to 2\,\textmu m.\cite{Schweizer2023} Desired thermal patterns were generated by phase-based wavefront modulation \cite{Vogel2015, Vogel2018} of the heating laser in combination with Fourier optics. The \textit{Cobolt Twist} laser source with a wavelength of \SI{457}{\nano \meter} is directed to a spatial light modulator, which imprints a spatial distribution of phase shifts. The intermediate image of the modified laser wavefront, visible after a lens, is focused on the sample via a microscope objective. Depending on the focal plane, we obtain a spot diameter between \SI{2}{\micro \meter} and \SI{9}{\micro \meter}. Spatial resolution is obtained by moving the thermal pattern over the sample surface. 

The magnon density spectrum $N(\omega)$ was measured by means of Brillouin light scattering (BLS) spectroscopy.\cite{sandercock1975, Buettner2000, Sebastian2015microBLS} This spectroscopy is based on the process of inelastic scattering of an incident photon by a magnon. The intensity of the inelastically scattered light is proportional to the density of magnons, whose frequency corresponds to the measured frequency shift of the photons. The probing laser source is a \textit{Coherent Verdi} laser operating at a wavelength of \SI{532}{\nano \meter}. The frequency of the scattered light was analyzed using the tandem multi-pass Fabry--P\'erot interferometer. To reduce the heating of the sample by the probing laser source we pulse the laser with an acousto-optic modulator. 

The described micro-BLS system was integrated with an optical heating system in one experimental setup as shown in \Fig{f:setup}(a). The setup was controlled using the \textit{thaTEC:OS} automation framework, and data evaluation was performed using Python libraries such as \textit{PyThat}\cite{PyThat} and \textit{xarray}\cite{Hoyer2017}.

First of all, we measured the relative number of thermal magnons $N(\omega)$ as a function of their frequency in the center of the hot spot.\cite{Schweizer2023} The resulting function $N(\omega)$ is presented in \Fig{f:setup}(b) in comparison to the dispersion curves of the fundamental dipole-exchange magnon mode shown in \Fig{f:setup}(c). $N(\omega)$ is proportional to the density of magnon states $D(\omega)\propto [d \omega (k)/ d k]^{-1} $ and has two clear peaks near $\omega_\|$ and $\omega_\perp$, where $D(\omega)$ formally goes to infinity. \looseness=-1

It can be seen that heating of the investigated region leads to a decrease in the frequency of the upper peak of the magnon density and an increase in the frequency of the peak at the bottom of the magnon spectrum. This behavior is in perfect agreement with our expectations.
{F}ollowing our calculations presented in \Fig{f:model}(b){, the shift of these frequencies by a few hundred MHz indicates}, {a} strong localized heating of the film reaching more than \SI{120}{\celsius}.

To reveal the effect of thermally created magnetic inhomogeneity on the magnon Bose--Einstein condensate (BEC), we equipped our setup with a specialized microwave circuit. It was used to create a dense magnon population, allowing us to reach the threshold of BEC formation. Therefore, a microstrip resonator with a width of \SI{50}{\micro \meter} and a length of \SI{3.5}{\milli \meter} is placed under the YIG sample tangentially magnetized by a 1510\,Oe field ($B \sb{SI units}$=\SI{151}{\milli\tesla}). Being driven by external microwave pulses, this microstrip induces a pumping magnetic field parallel to the external field $\B H \sb{ext}$ and along the equilibrium direction of the magnetization $\B M$. Thus, the geometry of parallel parametric pumping is fulfilled.\cite{Gurevich-Melkov1996} The energy transfer from the electromagnetic to the spin system happens in the form of a microwave photon with a wavenumber close to zero, which decays into two magnons with half the pumping frequency $\omega \sb{p} = \SI{12.705}{\giga \hertz}$ and opposite wave vectors.\cite{Melkov2000, Serga2012, Lvov2023} Due to four-magnon scattering processes, parametrically pumped magnons thermalize in the lower region of the spectrum and form a BEC at its bottom at $\omega_\|$ [see \Fig{f:setup}(c)] when the threshold density is reached.\cite{Demokritov2006, Pumping_parameters}

Figure\,\ref{f:freq_intens} shows the frequency-spatial magnon distribution on the longitudinal axis of the microstrip around the hot spot in the center of the pumping zone at different moments of time $t$. 
After the start of pumping [see \Fig{f:freq_intens}(a)], parametric magnons injected at $\omega_\perp = \omega\sb{p}/2=$~\SI{6.352}{\giga\hertz} move through the step-by-step Kolmogorov--Zakharov scattering cascade to $\omega_\|$.\cite{Lvov2024} Although, at this time, the expected decrease of $\omega_\perp$ is already noticeable in the hot region, nothing can be said about the behavior of $\omega_\|$ yet since the thermalizing magnons have not yet reached this frequency. 

{After some time}, [see \Figs{f:freq_intens}(b), (c)] the magnons fill the entire frequency region between $\omega_\perp$ and $\omega_\|$, concentrating at the bottom of the spectrum. Due to changes in the parametric pumping conditions in the hot spot, the density of gaseous magnons here is somewhat lower than in the surrounding areas. This is not the case for the near-bottom magnons, where the upward shift of $\omega_\|$ to about 80\,MHz becomes clearly visible. The frequency shift profile can be well approximated by a Gaussian curve with \SI{10}{\micro \meter} full width at half maximum. 

Of particular interest is the dynamics of the spatial distribution of near-bottom magnons. One can see from \Figs{f:freq_intens}(b)-(d) that during the pumping action and some time after its termination at $t=0$, the hot spot is surrounded by areas of increased magnon density. This phenomenon finds a natural interpretation in the dynamics of the magnon supercurrents flowing out of the region of increased frequency.\cite{Bozhko2019, Schweizer2022} The strength of the supercurrent depends on the phase gradient of the BEC wave function and, hence, on the BEC frequency gradient.\cite{Bozhko2016} Due to the decrease of this gradient with distance from the heating region, the supercurrent decreases, and magnons accumulate due to the ``bottle-neck'' effect when the inflow of quasiparticles exceeds their outflow. The decrease in the BEC density caused by its outflow from the hot region is compensated by the condensation of the parametrically overpopulated magnon gas. Such compensation ceases after the pumping is turned off, which leads to the formation of a deep BEC density dip in the hot spot [see \Figs{f:freq_intens}(d)-(f)]. The formation of this dip is further intensified by the growth of the supercurrent due to the increasing BEC coherence after the termination of the disturbing effect of pumping.\cite{Kreil2019, Noack2021} Soon, due to the decrease of supercurrents owing to the depletion of the magnon condensate in the region of maximum heating, the magnon density humps disappear as well [see \Figs{f:freq_intens}(e), (f)]. 
However, the spatial redistribution of the condensed magnons leads to the fact that during the entire time after pumping is turned off, the BEC density outside the hot spot remains higher than in the absence of heating. \looseness=-1

In summary, we see that an increase in the lower frequency limit of the magnon spectrum due to the influence of demagnetization fields in a locally heated region causes supercurrents to flow out of the heated area. At the same time, such an increase depends on the ratio of the diameter of this region and the film thickness, which opens up new opportunities for controlling magnon supercurrents in thermal landscapes. It can also be expected that in submicron-thick YIG films, where the lower frequency of the dipole-exchange magnon spectrum lies significantly above $\omega_\|$ and, therefore, strongly depends on the magnetization change, local heating can lead to an inversion of the supercurrent direction. 

    This study was funded by the Deutsche Forschungsgemeinschaft (DFG, German Research Foundation) -- TRR 173 -- 268565370 Spin+X (Project B04). V.S.L. was in part supported by NSF-BSF grant \# 2020765.



\bibliography{Supercurrent_direction} 

\begin{thebibliography}{43}
\bibitem [{Dem()}]{Demagnetization_factors}%
  \BibitemOpen
  \href@noop {} {}\bibinfo {note} {{Note that in the case of a sphere, $N_x=N_y=N_z=1/3$, whereas in an ellipsoid of revolution (spheroid) along $\widehat{\B z}$, $N_x=N_y$. In the case of a very elongated spheroid, $N_z$ tends to zero, and thus $N_x=N_y\to 1/2$. Similarly, for a very flattened spheroid, $N_z\to 1$, and $N_x=N_y= 0$}}\BibitemShut {NoStop}%
\bibitem [{Pum()}]{Pumping_parameters}%
  \BibitemOpen
  \href@noop {} {}\bibinfo {note} {{In our experiments, the pump pulse duration was \SI{500}{\nano \second}, and the maximal pump power was up to \SI{5}{\watt}. The parametric instability threshold was reached at a pump power $P\sb{p} = P\sb{p}\sp{thr} \simeq \SI{1.3}{\micro \watt}$, magnon accumulation at $\omega_\|$ began when $P\sb{p}$ was increased by 20\,dB, and the results presented in \Fig{f:freq_intens} were obtained at a pump power exceeding $P\sb{p}\sp{thr}$ by 36\,dB (\SI{5}{\watt}). Due to the relatively large size of the microstrip, and hence the area of magnon BEC formation, compared to the wavelength of condensed magnons $\lambda = 2\pi / k \simeq 1$\,\textmu m, we can talk about the formation of a spatially homogeneous BEC along the microstrip.}}\BibitemShut {Stop}%
\end{thebibliography}
 
\end{document}